\documentclass[12pt]{article}
\pdfoutput=1

\newcommand{\blind}{1}

\usepackage{graphicx}
\usepackage[title]{appendix}
\usepackage{inputenc}
\usepackage{ulem}
\usepackage{amsmath}
\usepackage{eurosym}
\usepackage{url}
\usepackage{placeins}
\usepackage{tabularx}
\usepackage{longtable}
\usepackage{adjustbox}

\usepackage[footnotesize,bf]{caption2}

\usepackage{bm}
\usepackage{float}
\usepackage{amsmath} 
\usepackage{amssymb} 
\usepackage{latexsym}
\usepackage{wasysym}
\usepackage[flushleft]{threeparttable}
\usepackage{color}
\usepackage{setspace}
\usepackage{lscape}
\usepackage{hhline}
\usepackage[round]{natbib}
\begin{document}
\def\spacingset#1{\renewcommand{\baselinestretch}%
	{#1}\small\normalsize} \spacingset{1}

\title{Unconventional Policies Effects on Stock Market Volatility: A MAP Approach} 
\date{}
\if1\blind
{
\author{Demetrio Lacava\footnote{University of Messina, email: dlacava@unime.it} \and Giampiero M. Gallo\footnote{Italian Court of Audits (Corte dei conti, Milan -- disclaimer) and New York University in Florence; email: giampiero.gallo@nyu.edu} \and  Edoardo Otranto\footnote{Corresponding author. University of Messina and CRENoS, Via dei Verdi 75, 98122 Messina, email: eotranto@unime.it}}
} \fi

\if1\blind
{
	\bigskip
	\bigskip
	\bigskip
	\medskip
} \fi

\maketitle              

\begin{abstract} Taking the European Central Bank unconventional policies as a reference, we suggest a class of Multiplicative Error Models (MEM) taylored to analyze the impact  such policies have on stock market volatility. The new set of models, called MEM with Asymmetry and Policy effects (MAP), keeps the base volatility dynamics separate from a component reproducing policy effects, with an increase in volatility on announcement days and a decrease unfolding implementation effects. When applied to four Eurozone markets, a Model Confidence Set approach finds a significant improvement of the forecasting power of the proxy after the Expanded Asset Purchase Programme implementation; a multi--step ahead forecasting exercise estimates the duration of the effect, and, by shocking the policy variable, we are able to quantify the reduction in volatility which is more marked for debt--troubled countries.\\


\end{abstract}
\vfill
\newpage
\spacingset{1.45} 

\section{Introduction}

During the Great Recession, with policy interest rates reaching the zero lower bound, many central banks, the Federal Reserve (FED), the European Central Bank (ECB) and the Bank of England (BoE) in particular, adopted unconventional actions in order to stimulate the real economy. Among unconventional monetary policies, one finds the central bank balance sheet expansion -- generally through Asset Purchase Programs (APP) -- which affects the real economy by reducing long--term yields and the credit spread, and by  modifying inflation rate expectations during periods in which the liquidity trap makes the conventional policy no longer effective. Since 2009, the ECB resorted to so--called Quantitative Easing (QE) in order to inject liquidity into the financial system. Geared toward preventing a systemic collapse due to the deterioration of credit ratings, but also to an adverse market attitude toward  debt--ridden economies, these policies have been credited with successfully reducing market uncertainty and increasing Eurozone stability. 

For the bond markets, the effects of monetary policy actions are studied in two directions: an increase in  prices due to increased direct demand (and hence a reduction in yields) and a lower volatility (because of the increased stability brought to the system). Among the empirical analyses we cite \cite{Eser:Schwaab:2016} and \cite{DeSantis:2020}; \cite{Ghysels:Idier:Manganelli:Vergote:2017} use intra--daily high frequency data to document an asymmetric effect due to  the presence of Security Market Programme (SMP) purchases in a GARCH(1,1). They find that unconventional monetary policies by ECB had a highly statistically significant dampening impact on yield volatility for most countries and maturities.\footnote{Rather than lengthening the citations of this rich literature in the bond markets, as they address issues and transmission channels which are not the main focus of this paper, we prefer to summarize some of them in a table in the supplemental material.}

As per the stock market, the directional effect is less pronounced: 
event studies pinpoint to single positive effects on stock prices \citep{Haitsma:Unalmis:DeHaan:2016}; 
the transmission of monetary policy to the stock market can be traced to the portfolio-rebalancing channel  \citep{Georgiadis:Graab:2016} in redetermining asset allocations as an indirect result of bond price movements; inflation expectations have been less of a concern for at least the past fifteen years, while  the confidence channel  mitigates general uncertainty with what we consider a price direction--neutral effect. 

The focus of this paper is the impact of the QE on the volatility of stock markets: while the methodology of choice in other papers is a GARCH model \citep{Engle:1982,Bollerslev:1986}, in which some policy related variable (usually one or more dummies) is inserted in the conditional variance equation,  
in what follows, we exploit the established merits of working with a realized measure of volatility  \citep{Andersen:Bollerslev:Christoffersen:Diebold:2006}. 
Being a positive--valued process, this measure is suitable to be modeled for forecasting purposes as a Multiplicative Error Model, MEM \citep{Engle:2002,Engle:Gallo:2006}, in which volatility is the product of a time-varying factor (following a GARCH--type process) and a positive random variable ensuring positiveness without resorting to logs. Within the MEM class, \cite{Brownlees:Cipollini:Gallo:2012} propose a model - the Composite AMEM (ACM) - in which the conditional expected volatility is the sum of a short-- and a long--run component. This model suggests a way to separate two volatility components, one representing the usual autoregressive dynamics of volatility and the other where we gather both the announcement and the implementation effects of unconventional policies. As a by--product of estimating the latent factors, we can calculate the share of the policy--related component on total expected volatility. To extend this model, labeled MEM with Asymmetry and Policy effects (MAP), we consider the case in which the two latent factors, rather than combining additively, are modified to be a product of each other (with two specifications adopted).  

The empirical application analyzes the impact of unconventional monetary policies by the ECB on stock market volatility, taking four Eurozone countries (France, Germany, Italy and Spain) as our leading case, considering the latter two as representative of debt--burdened markets. We proxy for unconventional policies by using the ratio between the securities purchased by the ECB for unconventional policy purposes and the ECB total asset \citep[similarly to][]{dAmico:English:LopezSalido:Nelson:2012,Voutsinas:Werner:2011}. Our main results show that announcements cause an immediate peak in volatility, whereas the implementation of unconventional policies has a mitigating effect along time; moreover, including these variables in the model, improves the out--of--sample forecasting, in particular after the EAPP announcements. 

The paper is organized as follows. Section \ref{sec:unconventional} describes data as well as the stylized facts on volatility with respect to the unconventional policies implementation. Section \ref{sec:model} introduces the novel MEM--based models employed in our empirical analysis, in Section \ref{sec:empirical}. We discuss model estimation and inference in Sub--section \ref{sec:estim}, model comparisons including an analysis of which models enter Model Confidence Set in various subsamples (in \ref{sec:MCS})  and a multi--step forecasting exercise to determine the estimated duration of the effects and the volatility response to a shock to the policy variable (in \ref{sec:converge}). Finally, Section \ref{sec:concl} concludes with some remarks.

\section{Proxies for unconventional monetary policies}
\label{sec:unconventional}

In investigating the impact of unconventional monetary policies by ECB we consider two different variables, which refer to announcement and implementation effects on volatility, respectively. The former is  measured by means of a dummy variable taking value of 1 on days in which ECB releases communications regarding a monetary policy decision\footnote{\url{https://www.ecb.europa.eu/press/pr/activities/mopo/html/index.en.html}}. To proxy for the implementation effect we use the amount of securities held by ECB as a fraction of total asset\footnote{Data are taken from the ECB website and Datastream.}  \citep[cf.][]{dAmico:English:LopezSalido:Nelson:2012,Voutsinas:Werner:2011}.

Our analysis is based on a dataset consisting of 2686 daily observations of annualized realized kernel volatility\footnote{Data from \url{https://realized.oxford-man.ox.ac.uk/data/download}} (RV hereafter) which is a robust estimator of the volatility, in particular with respect to microstructure noise \citep{BarndorffNielsen:Hansen:Lunde:Shephard:2008}. The analysis relates to four Eurozone market indices (CAC40 for France, DAX30 for Germany, FTSE MIB for Italy and IBEX35 for Spain -- referred to by country in what follows) for the periods between June 1, 2009 and December 31, 2019.

We analyze this dataset by a simple visual inspection of the graphs, highlighting the main announcements in the time span considered, and by analyzing some stylized facts that justify the model proposed in Section \ref{sec:model}.

\subsection{Graphical Analysis of Announcements}
\label{sec:vis_insp}
\begin{figure}[t]
	\centering
	\includegraphics[scale=0.68]{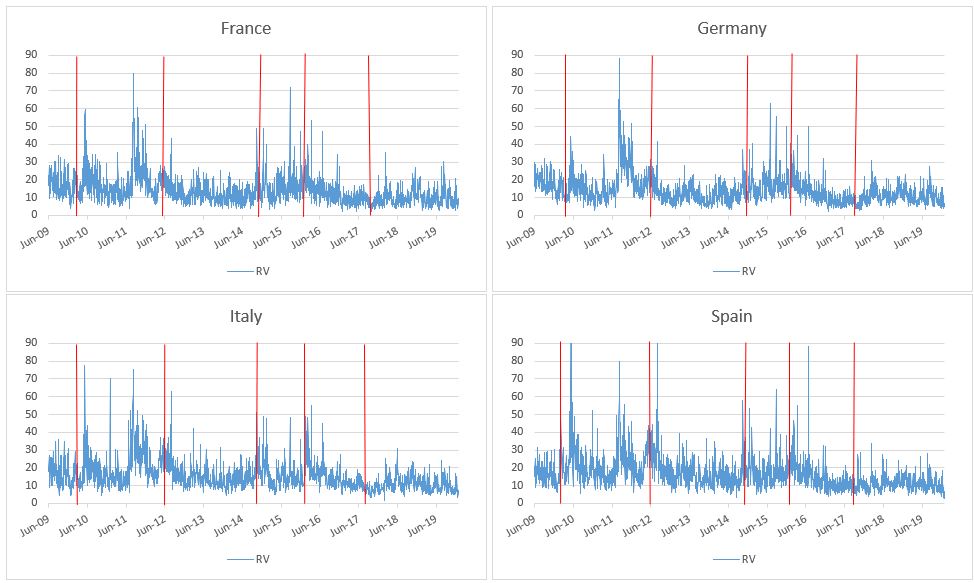}
	\caption{France, Germany, Italy and Spain RV. Sample period: June 1, 2009 to December 31, 2019. Number of observations:
		2686. The vertical lines represent relevant events for policy actions (see text).} \label{fig:RVseries}
\end{figure}

Figure \ref{fig:RVseries} shows the evolution of our series. All  series seem to follow a similar pattern, with a period of low volatility in the first year of the sample. However, while the RV series seem not to respond to the unconventional monetary policies established by ECB at the beginning of the crisis,\footnote{These are the 12-month Longer Term Refinancing Operations program  - the LTRO - and the Covered Bond Purchase Programme - the CBPP - which aimed at containing the liquidity crisis and the consequent credit crunch for the Eurozone.} starting from May 2010, in all the series, one can notice how the RV reacted to some important events (reported as vertical lines in Figure \ref{fig:RVseries}), such as:
\begin{itemize}
\item the SMP (Security Market Programme) announcement on May 10th, 2010. By means of purchases of government bond in the secondary market, the ECB aimed to control the increase in the credit spread and to restore the proper functioning of monetary policy transmission channels. While the RV jumped on the day when the SMP was announced, the subsequent implementation of this program had a dampening effect on volatility.
\item the ``Whatever it takes" declaration by Mario Draghi on July 26th, 2012, which served to reassure investors regarding the emerging denomination risk. Through this declaration, the ECB announced the Outright Monetary Transaction (OMT), which replaced the SMP successfully, depressing volatility until the end of 2014.
\item the EAPP (Expanded Asset Purchase Programme) announcement on January 22nd, 2015. It was established mainly to improve monetary policy transmission mechanisms as well as to adjust the inflation rate toward the target level of 2\%. It refers to a series of unconventional measures such as the Assed Backed Securities Purchase Programme (ABSPP), the CBPPs and the Corporate and Public Sector Purchase Programme (CSPP and PSPP, respectively), through which ECB
conducted monthly securities purchases.
\item March 10th, 2016. The amount of securities purchased within the EAPP passed from the initial level of \EUR 60 billion to \EUR 80 billion per month, causing a downward trend in the RV series.
\item October 26th, 2017. Volatility increases after the announcement through which ECB communicated the cut in the monthly purchases, which were reduced to \EUR 15 billion. In contrast to the previous announcement, it caused an increasing trend in all the considered markets.
\end{itemize}
A simple visual inspection shows the decrease in volatility in correspondence of these events; moreover, we notice also an effect caused by the amount purchased by ECB. We aim to quantify this effect and its weight on the level of volatility.

\subsection{Stylized facts}
\label{sec:sty_facts}

In order to analyze the effects of unconventional monetary policy on volatility, one would have to build a structural model detailing the channels of transmission onto financial markets, isolating the nature of the innovations. What is preferred in empirical investigations is a reduced--form approach in which the net effect of certain variables is examined without resorting to a structural explanation, also not to enter in the subjective choice about which theory to adopt \citep{Ghysels:Idier:Manganelli:Vergote:2017}. We adhere to this strategy, starting from some stylized facts which may serve as the basis for making time series modeling choices later in the paper. We will not delve on the well--known persistence features of realized volatility and its asymmetric behavior relative to upward and downward movements in a market. We rather suggest a few results about the relationship between volatility and policy variables. 
 
The first stylized fact is about higher volatility on the days of announcements. Not being able to repeat a visual inspection of our set of 144 ECB announcements in the time span considered (as we did in  Figure \ref{fig:RVseries}), we suggest to see whether there is a pattern in terms of volatility behavior before and after an announcement, by calculating the average realized volatility 5 days before and 5 days after the announcements;\footnote{The only exception relates to first announcement occurred at $t=4$, so that the average in that case involves only 3 terms before and after.} Table \ref{tab:average} shows how these averages compare to the volatility on the day of the announcement in relative percentage terms. As all the variations are negative, the announcement days mark, on average, a peak of volatility, with a subsequent reduction after the announcement. 

\begin{table}[t]
	\begin{center}
		\caption{Percentage variation of the average of volatility (5 days before and after the announcement date) with respect to the volatility of the announcement day for four volatility time series.}\label{tab:average}
		\begin{tabular}{lrrrr}
		&France&Germany&Italy&Spain\\ \hline	
		before&-4,40&	-5,92&	-9,79&	-8,56\\
		after&-5,52&	-5,39&	-7,39&	-5,02\\ \hline
	\end{tabular}
\end{center}
\end{table}
			
A second empirical regularity is about the direction of the causal relationship (in a Granger's sense) between the realized volatility and the unconventional policy proxy (allowing also for a dummy on announcement days). The non--causality tests\footnote{In view of the stationarity of the realized volatility and of the nonstationarity of the policy variable, we adopted the procedure suggested by \cite{Toda:Yamamoto:1995} which envisages a VAR(6) in levels with five lags on either side to be cross--tested and an extra one  (not to be tested).} seem to support the idea that the unconventional policy affects the realized volatility, but not for the reverse. 

A third issue is whether there is a systematic difference between average returns on announcement and non announcement days, simply assessed by regressing  returns standardized by the realized volatility on the dummy for announcement days. Since the corresponding coefficient is never significant, we proceed with the empirical regularity that there is no systematic impact of the announcements on returns. 

\section{Policy--related volatility components}
\label{sec:model}

Tha availability of high--frequency data has given rise to a stream of literature on realized variance measures \citep{Andersen:Bollerslev:Christoffersen:Diebold:2006} with models based on the features of volatility traditionally investigated in the GARCH framework \citep{Engle:1982,Bollerslev:1986}. Among these, a prominent place is taken by Multiplicative Error Models (MEM) as 
defined by \cite{Engle:2002} and successively revised by \cite{Engle:Gallo:2006} to allow for asymmetric effects (AMEM).

The model we propose is an extension of the Composite AMEM (ACM) \citep[][to which we refer for details]{Brownlees:Cipollini:Gallo:2012} in which the conditional expected volatility is the sum of a short-- and a long--run component.\footnote{Another model treating volatility components additively is the Spillover AMEM (SAMEM) by  \cite{Otranto:2015}, which captures volatility spillovers among markets in a univariate framework.}  In this ACM we insert the effect of unconventional policies as a latent factor, explicitly affecting the dynamics of the volatility. More precisely, our model is based on the decomposition of the volatility level in the sum of two components, representing the base volatility component ($\varsigma$) and what can be related to the unconventional policies ($\xi$), respectively, allowing us to quantify its relative effect on the overall level of volatility.

Thus, our composite MEM with Asymmetry and Policy effects (MAP) consists of four equations:
\begin{equation}
\begin{array}{l}
 RV_{t}=\mu _{t}\epsilon_{t}, \quad \epsilon _{t}|\Psi
_{t-1}\sim Gamma(\vartheta ,\frac{1}{\vartheta })\\
\mu _{t}=\varsigma _{t}+\xi _{t}\\
\varsigma _{t}=\omega +\alpha RV_{t-1}+\beta \varsigma
_{t-1}+\gamma D_{t-1}RV_{t-1}\\
\xi _{t}=\delta (E\left(x_{t}|\Psi_{t-1}\right)-x^\star)+ \varphi (\Delta_{t}-\Delta^ \star)+ \psi \xi _{t-1}
\end{array} \label{eq:ACM}
\end{equation}
where $\Psi _{t}$ is the information set available at time $t$. Following \cite{Engle:Gallo:2006}, we consider the error term $\epsilon_t$ as Gamma distributed,  in view of its flexible dependence on just one parameter, $\vartheta$ and of its good empirical performance. From these assumptions, we get not only a time--varying conditional mean $E\left(RV_t|\Psi_{t-1}\right)=\mu_t$, but also a time--varying conditional variance (volatility of volatility) $Var\left(RV_t|\Psi_{t-1}\right)=\mu_t^2/\vartheta$; as $\varsigma _{t}$ follows a GARCH--type dynamics, the model is suitable to capture possible clustering in volatility. In our model (\ref{eq:ACM}), $D_{t-1}$ represents a dummy variable taking value of 1 if the return of the asset (index) at time $t-1$ is negative, 0 otherwise.

The additional component $\xi _{t}$ is an AR(1) involving $x_t$ and $\Delta_t$ as predetermined variables (in view of the stylized facts), representing, respectively, the implementation and announcement effect; finally, $x^\star$ is a constant, e.g. the unconditional mean or an initial value (to accomodate a random walk process). Similar remarks apply to  $\Delta^\star$ and $\Delta_t$; this structure, under the hypothesis of stationarity, ensures a zero unconditional mean for $\xi_t$. As for the timing of these predetermined variables, since the days of monetary policy announcements are put on the calendar well in advance, we can consider the announcement variable, $\Delta_{t}$, at its contemporaneous time. By contrast, the proxy representing the implementation effect has to enter in expected terms: for simplicity, given its behavior assessed as a random walk, we adopt the lagged value as an expectation. Thus, by also replacing $x^\star$ and $\Delta^\star$ with the respective sample means $\bar{x}$ and $\bar{\Delta}$, the last equation in model (\ref{eq:ACM})  for operational purposes will become:  
$ \xi _{t}=\delta (x_{t-1}-\bar{x})+ \varphi (\Delta_{t}-\bar{\Delta})+ \psi \xi _{t-1}$. 

Note that in Equation (\ref{eq:ACM}), when we set $\psi=\beta$, we get the X--MAP specification for $\mu_t$ (which corresponds to an AMEM with an exogenous regressor):
\begin{equation}
	\mu _{t}=\omega +\alpha RV_{t-1}+\beta \mu _{t-1}+\gamma
	D_{t-1}RV_{t-1}+ \delta (E\left(x_{t}|\Psi_{t-1}\right)-x^ \star)+ \varphi (\Delta_{t}-\Delta^ \star),
	\label{eq:X-MAP}
\end{equation}
which, in turn, nests the classical AMEM  upon imposing  $\delta=\varphi=0$.

For stationarity in the MAP, it is required that both components are stationary in covariance, that is $(\alpha + \beta+ \frac{\gamma}{2})<1$ and $\psi<1$; positiveness, instead, requires  that ($\varsigma_t+\xi_t>0)$ for each $t$, which could be ensured even though $\delta$ is negative, as we expect.

Finally, as argued by \cite{Engle:Lee:1999}, the coefficient $\psi$ in $\xi_t$ process is required to be less than $\beta$ to ensure the identification of the model. Since one of the main ECB aims by means of unconventional policies is to stabilize financial markets in the short run, we expect an immediate effect of this kind of policy in reducing stock market volatility. In other words, in our model, the part of volatility depending directly on unconventional policies represents the short run component of realized volatility as well as the proper volatility dynamics represents the long--run component. From this assumption, we expect the long--run component to have a higher persistence than the short one, that is $0<\psi<\beta<1$, as an identification condition.

It is important to underline that $\xi_t$ is an unobservable signal, with its own dynamics, which can be jointly estimated so as to quantify and plot the separate effect of the unconventional ECB actions on the volatility $RV_t$. As a by--product, we can also derive the share of this component relative to the overall level of volatility, that is  $1 - \frac{\varsigma_t}{\mu_t}=\frac{\xi_t}{\mu_t}$.

As shown by \cite{Engle:2002} for the MEM case, quasi maximum likelihood maximization implies that the estimators of the unknown coefficients in model (\ref{eq:ACM}) are consistent and asymptotically normal, under an assumption of correct specification of the conditional mean equation, irrespective of the distributional assumption, as discussed by \cite{Engle:Gallo:2006}. When the scale parameter $\theta$ is unknown, robust standard errors will shield against the actual shape of the Gamma distribution.

Whether the policy impact on the general level of volatility should enter in an additive way, as in model (\ref{eq:ACM}), or multiplicatively is always open to question.\footnote{Within the MEM class, a multiplicative component MEM is suggested by \cite{Brownlees:Cipollini:Gallo:2011}, adapting a GARCH approach by \cite{Engle:Sokalska:2012}. } In what follows we discuss two different specifications of the multiplicative version of MAP which ensure the compliance with the constraint that the unconditional mean of $\xi_t$ is equal to one so that the model is identified.

In the first specification we allow $\xi_t$ to impact on $RV_t$ through a logistic function - and by means of the delta method. The model, called Logistic-MAP (L-MAP), is specified as in model (\ref{eq:L-MAP}) 
\begin{equation}
\begin{array}{l}
 RV_{t}=\mu _{t}\epsilon_{t}, \quad \epsilon _{t}|\Psi
_{t-1}\sim Gamma(\vartheta ,\frac{1}{\vartheta })\\
\mu_{t}=2 \varsigma _{t}(\frac{exp(\xi_t)}{1+exp(\xi _{t})})\\
\varsigma _{t}=\omega +\alpha RV_{t-1}+\beta \varsigma
_{t-1}+\gamma D_{t-1}RV_{t-1}\\
\xi_{t}=\delta (E\left(x_{t}|\Psi_{t-1}\right)-x^ \star)+ \varphi (\Delta_{t}-\Delta^ \star)+ \psi \xi _{t-1}
\end{array} \label{eq:L-MAP}
\end{equation}

Alternatively, we can add a constant term in the $\xi_{t}$ equation, providing a Plain-MAP (P-MAP) specification, namely:
\begin{equation}
\begin{array}{l}
 RV_{t}=\mu _{t}\epsilon_{t}, \quad \epsilon _{t}|\Psi
_{t-1}\sim Gamma(\vartheta ,\frac{1}{\vartheta })\\
\mu _{t}=\varsigma _{t}\xi _{t}\\
\varsigma _{t}=\omega +\alpha RV_{t-1}+\beta \varsigma
_{t-1}+\gamma D_{t-1}RV_{t-1}\\
\xi _{t}=(1-\psi)+\delta( E\left(x_{t}|\Psi_{t-1}\right)-x^\star)+ \varphi (\Delta_{t}-\Delta^\star)+ \psi \xi _{t-1}
\end{array} \label{eq:P-MAP}
\end{equation}


A way to characterize how the suggested models differ in terms of how the policy proxy impacts expected volatility is by calculating the marginal effects of a change in $x_{t-1}$ (respectively, $\Delta_t$) on $\mu_{t+ \tau}$ ($\tau$-steps ahead), as done in Table \ref{tab:marginal_effects_formula}.\footnote{The derivation of the formulas is available as supplemental material.}

\begin{table}[t]
	\begin{center}
		\caption{Marginal effects of policy variables $x_{t-1}$ and $\Delta_t$ on $\mu_{t+\tau}$, $\tau\geq0$.}
		\begin{tabular}{cc}
			Model                             & Marginal effect on $\mu_{t+\tau}$\\ [2mm]
			\hline \\[-2mm]
			
			MAP& $\kappa \psi^\tau$  \\  [2mm]
			X--MAP&  $\kappa \beta^\tau$   \\ [2mm]
			L-MAP & $2\varsigma_{t+\tau} \kappa \psi^\tau \frac{exp(\xi_{t+\tau})}{(1+exp(\xi_{t+\tau}))^2}$ \\ 
			[2mm]
			P-MAP   &$\kappa \psi^\tau \varsigma_{t+\tau}$ \\ [2mm]
			\hline 
		\end{tabular}\label{tab:marginal_effects_formula}
	\end{center}
	\begin{tablenotes}
		\footnotesize
		\item  \textit{Note:} $\kappa=\delta$ for the marginal effects of the implementation variable $x_{t-1}$;\\
		$~\ ~\ \qquad \kappa=\varphi$ for the marginal effects of the announcement variable $\Delta_{t}$.
	\end{tablenotes}
\end{table}

\section{Empirical application}
\label{sec:empirical}

\subsection{Estimation results}
\label{sec:estim}
The goal of this section is to show  how different models for the realized volatility series of the four indices take the policy variables into consideration and how the effects are exerted on the level of volatility in different specifications.

Parameter estimation of the different models are shown in Table \ref{tab:results}. Coefficients are highly significant in all cases with a volatility persistence ($\alpha+\beta+\frac{\gamma}{2}$) which decreases by about 2\% passing from the AMEM to our more complex MAP. Moreover, the impact of the more recent observed values, measured by the $\alpha$ and $\gamma$ coefficients, is generally higher in the AMEM. This indicates a lower influence of current shocks on volatility projections, in line with the expected calming effect that the unconventional monetary policies have on market volatility.

Diagnostics on the standardized residuals is given by the Ljung-Box statistics  for lags 1, 5 and 10 (at the bottom of the table): we see how the AMEM is able to capture the persistence in the volatilities (we fail to reject the null of no serial correlation at 1\% significance level, with the exception of Germany at the highest lag, as well as Spain at lag 5). However, our additive model (the MAP) seems to have a better statistical performance in this respect, especially in the case of Spain, where we never reject the null of no autocorrelation at 1\% level.

Model performance in terms of estimation results can be compared in Table \ref{tab:comparison}, where we report the information criteria (AIC and BIC) and two loss functions  \citep[Mean Square Error --MSE -- and Quasi-Likelihood -- QLike, consistent in the sense of ][]{Patton:2011} to evaluate the fitting capabilities of the models. For all countries, the best performing  model (the one in bold) is the L-MAP. The only exception is Spain, where the L-MAP has better performance in terms of the information criteria, even if the P-MAP is marginally better in terms of MSE.

	\begin{table}[t]
	\caption{Models estimation results. Sample: June 1, 2009 -- December 31, 2019.}
	\scriptsize{
		\begin{tabular}{lrrrrr|rrrrr}
			& \multicolumn{5}{c}{\textbf{France}} & \multicolumn{5}{c}{\textbf{Germany}}\\ [1mm]
			& AMEM & X--MAP & MAP & L--MAP & P--MAP & AMEM & X--MAP & MAP & L--MAP & P--MAP\\
			\hhline{===========}
			\multicolumn{11}{c}{Coefficient estimates (robust s.e. in parentheses)}\\
			$\omega$ & 0.857  & 1.136  & 1.056  & 1.011  & 1.025  & 0.957  & 1.081  & 1.034  & 1.026  & 1.034  \\
			& (0.046) & (0.121) & (0.065) &   (0.007) &   (0.053) & (0.017)  & (0.052) & (0.016) & (0.059) &   (0.519) \\[1mm]
			$\alpha$ & 0.171  & 0.165  & 0.154  & 0.151  & 0.153  & 0.193  & 0.191  & 0.183  & 0.182  & 0.183  \\
			& (0.017) & (0.020) & (0.015) &   (0.015) &   (0.017) & (0.010)  & (0.011) & (0.014) & (0.012) &   (0.055) \\[1mm]
			$\beta$  & 0.708  & 0.689  & 0.707  & 0.712  & 0.709  & 0.692  & 0.684  & 0.696  & 0.696  & 0.694  \\
			& (0.019) & (0.029) & (0.015) &   (0.014) &   (0.018) & (0.011) & (0.016) & (0.013) & (0.014) &   (0.094) \\[1mm]
			$\gamma$ & 0.113  & 0.120  & 0.117  & 0.119  & 0.119  & 0.087  & 0.090  & 0.089  & 0.090  & 0.090  \\
			& (0.008) & (0.009) & (0.011) &   (0.010) &   (0.012) & (0.006) & (0.011) & (0.008) & (0.008) &   (0.008) \\[1mm]
			$\delta$ &       & -0.636  & -1.836  & -0.297  & -0.161  &       & -0.454  & -1.328  & -0.219  & -0.115  \\
			&       & (0.075) & (0.326) &   (0.057) &   (0.029) &       & (0.035) & (0.283) & (0.052) &   (0.057) \\[1mm]
			$\varphi$ &       & 1.297  & 2.817  & 0.464  & 0.231  &       & 1.067  & 2.317  & 0.378  & 0.188  \\
			&       & (0.381) & (0.539) &   (0.093) &   (0.045) &       & (0.311) & (0.470) & (0.073) &   (0.073) \\[1mm]
			$\psi$   &       &       & 0.111  & 0.194  & 0.134 &       &       & 0.098  & 0.175  & 0.138  \\
			&       &       &   (0.060) &   (0.081) & (0.083) &       &       & (0.077) & (0.097) &   (0.204) \\[1mm]
			$\theta$ & 7.559  & 7.728  & 7.817  & 7.827  & 7.820  & 9.460  & 9.610  & 9.700  & 9.719  & 9.714  \\
			& (0.222) & (0.228) & (0.231) &   (0.230) &   (0.231) & (0.290) & (0.298) & (0.301) & (0.301) &   (0.301) \\[2mm]
			Loglik & -7825.1 & -7793.9 & -7778.1 & -7776.2 & -7777.4 
			& -7592.5 & -7570.6 & -7557.8 & -7555.0 & -7555.7 \\ [2mm]
			\multicolumn{11}{c}{p--values for Ljung-Box statistics} \\
			lag 1 & 0.155 & 0.258 & 0.132 & 0.108 & 0.125 & 0.171 & 0.262 & 0.184 & 0.173 & 0.194 \\
			lag 5 & 0.111 & 0.167 & 0.215 & 0.192 & 0.207 & 0.028  & 0.063  & 0.056  & 0.051  & 0.054  \\
			lag 10 & 0.115 & 0.137 & 0.210  & 0.201 & 0.201 & 0.002  & 0.004  & 0.003  & 0.003  & 0.003  \\
			\multicolumn{11}{c}{} \\
			& \multicolumn{5}{c}{\textbf{Italy}} & \multicolumn{5}{c}{\textbf{Spain}}\\ [1mm]
			& AMEM & X--MAP & MAP & L--MAP & P--MAP & AMEM & X--MAP & MAP & L--MAP & P--MAP\\
			\hhline{===========}
			\multicolumn{11}{c}{Coefficient estimates (robust s.e. in parentheses)}\\
			$\omega$ & 1.198  & 1.708  & 1.533  & 1.480  & 1.490  & 1.055  & 1.533  & 1.438  & 1.410  & 1.426  \\
			& (0.073) & (0.221) & (0.260) & (0.239) & (0.146) & (0.075) & (0.110) & (0.135) & (0.621) &   (0.132) \\[1mm]
			$\alpha$ & 0.268  & 0.274  & 0.249  & 0.248  & 0.250  & 0.224  & 0.205  & 0.197  & 0.193  & 0.195  \\
			& (0.018) & (0.032) & (0.033) & (0.042) & (0.022) & (0.013) & (0.014) & (0.024) & (0.060) &   (0.023) \\[1mm]
			$\beta$  & 0.608  & 0.568  & 0.604  & 0.605  & 0.603  & 0.671  & 0.656  & 0.671  & 0.674  & 0.671  \\
			& (0.023) & (0.043)  & (0.043) & (0.049) & (0.030) & (0.007) & (0.020)  & (0.027) & (0.091) &   (0.028) \\[1mm]
			$\gamma$ & 0.084  & 0.086  & 0.086  & 0.089  & 0.088  & 0.078  & 0.087  & 0.085  & 0.086  & 0.086  \\
			& (0.005) & (0.011) & (0.019) & (0.022) & (0.008) & (0.022) & (0.007)  & (0.015) & (0.017) &   (0.009) \\[1mm]
			$\delta$ &       & -1.074  & -2.354  & -0.343  & -0.178  &       & -1.025  & -2.823  & -0.389  & -0.207  \\
			&       & (0.161) & (0.519) & (0.048) & (0.031) &       & (0.086)  & (0.528) & (0.133) &   (0.037) \\[1mm]
			$\varphi$ &       & 2.059  & 3.448  & 0.518  & 0.254  &       & 1.964  & 3.428  & 0.501  & 0.247  \\
			&       & (0.471) & (0.569) & (0.084) & (0.038) &       & (0.361)  & (0.671) & (0.087) &   (0.042) \\[1mm]
			$\psi$   &       &       & 0.051  & 0.089  & 0.058 &       &       & 0.058  & 0.123 & 0.072 \\
			&       &       & (0.083) & (0.055) & (0.073) &       &       & (0.090) & (0.298) & (0.112) \\[1mm]
			$\theta$ & 10.593  & 11.030  & 11.183  & 11.185  & 11.181  & 9.149  & 9.487  & 9.580  & 9.626  & 9.623  \\
			&  (0.423) & (0.427) & (0.436) & (0.437) & (0.436) &   (0.317) & (0.338)  & (0.349) & (0.355) &   (0.352) \\[2mm]
			
			Loglik & -7721.3 & -7665.3 & -7646.3 & -7646.1 & -7646.5 
			& -8117.0 & -8066.5 & -8052.9 & -8046.3 & -8046.7 \\  [2mm]
			
			\multicolumn{11}{c}{p--values for Ljung-Box statistics} \\
			lag 1 & 0.299 & 0.691 & 0.300   & 0.287 & 0.324 & 0.010 & 0.025  & 0.011  & 0.006  & 0.008  \\
			lag 5 & 0.497 & 0.721 & 0.800  & 0.805 & 0.824 & 0.009  & 0.015  & 0.019  & 0.012  & 0.015  \\
			lag 10 & 0.336 & 0.472 & 0.647 & 0.625 & 0.624 & 0.047  & 0.086  & 0.119 & 0.089  & 0.100 \\
			\hline 
		\end{tabular}\label{tab:results}
	}
\end{table}
\FloatBarrier

\begin{table}[h]
	\centering
	\caption{Model comparisons via Information Criteria (AIC and BIC) and forecasting capability (MSE and MAE) -- Sample period: June 1, 2009 - December 31, 2019. Best model in bold. }
	\vspace{2mm}
	\footnotesize{
		\begin{tabular}{lrrrrr}
			\multicolumn{1}{c}{} & \textbf{AMEM} & \textbf{X-MAP} & \textbf{MAP} & \textbf{L-MAP} & \textbf{P-MAP} \\
			\multicolumn{6}{c}{\textbf{France}} \\
			\hline
			AIC   & 5.830  & 5.809 & 5.797 & \textbf{5.796} & 5.797 \\
			BIC   & 5.841 & 5.824 & 5.815 & \textbf{5.814} & 5.815 \\
			MSE   & 29.531 & 29.117 & 28.844 & \textbf{28.627} & 28.639 \\
			QLike   & 0.068 & 0.066 & \textbf{0.065} & \textbf{0.065} & \textbf{0.065} \\
			\multicolumn{1}{c}{} &       &       &       &       &  \\
			\multicolumn{6}{c}{\textbf{Germany}} \\
			\hline
			AIC   & 5.657 & 5.642 & 5.633 & \textbf{5.631} & 5.632 \\
			BIC   & 5.668 & 5.658 & 5.651 & \textbf{5.649} & \textbf{5.649} \\
			MSE   & 23.551 & 23.269 & 23.096 & \textbf{22.93} & 22.939 \\
			QLike   & 0.054 & 0.053 & \textbf{0.052} & \textbf{0.052} & \textbf{0.052} \\ 
			\multicolumn{1}{r}{} &       &       &       &       &  \\
			\multicolumn{6}{c}{\textbf{Italy}} \\
			\hline
			AIC   & 5.753 & 5.713 & \textbf{5.699} & \textbf{5.699} & \textbf{5.699} \\
			BIC   & 5.764 & 5.728 & \textbf{5.717} & \textbf{5.717} & \textbf{5.717} \\
			MSE   & 28.171 & 27.634 & 27.209 & \textbf{27.191} & 27.203 \\
			QLike   & 0.048 & 0.046  & \textbf{0.045} & \textbf{0.045} & \textbf{0.045}\\
			\multicolumn{1}{r}{} &       &       &       &       &  \\
			\multicolumn{6}{c}{\textbf{Spain}} \\
			\hline
			AIC   & 6.048 & 6.011 & 6.002 & \textbf{5.997} & \textbf{5.997} \\
			BIC   & 6.059 & 6.027 & 6.02  & \textbf{6.015} & \textbf{6.015} \\
			MSE   & 42.048 & 40.918 & 40.462 & 39.461 & \textbf{39.439} \\
			QLike   & 0.056 & 0.054 & \textbf{0.053} & \textbf{0.053} & \textbf{0.053} \\
			\hline
		\end{tabular}%
	}
	\label{tab:comparison}
\end{table}

For what concerns the unconventional policy proxies, coefficients are significant at 1\% level and they enter the model with the expected sign. According to other results in this stream of research \citep[see, for example,][]{Bomfim:2003, Chan:Gray:2018, Shogbuyi:Steeley:2017}, the coefficient $\varphi$ of the dummy variable is positive, meaning that there is an immediate reaction in the market on announcement days, in line with our stylized facts. Conversely, $\delta$ is negative, therefore the unconventional policies implementation successfully reduces stock market volatility, as expected.

\begin{figure}[t]
	\centering
	\includegraphics[scale=0.6]{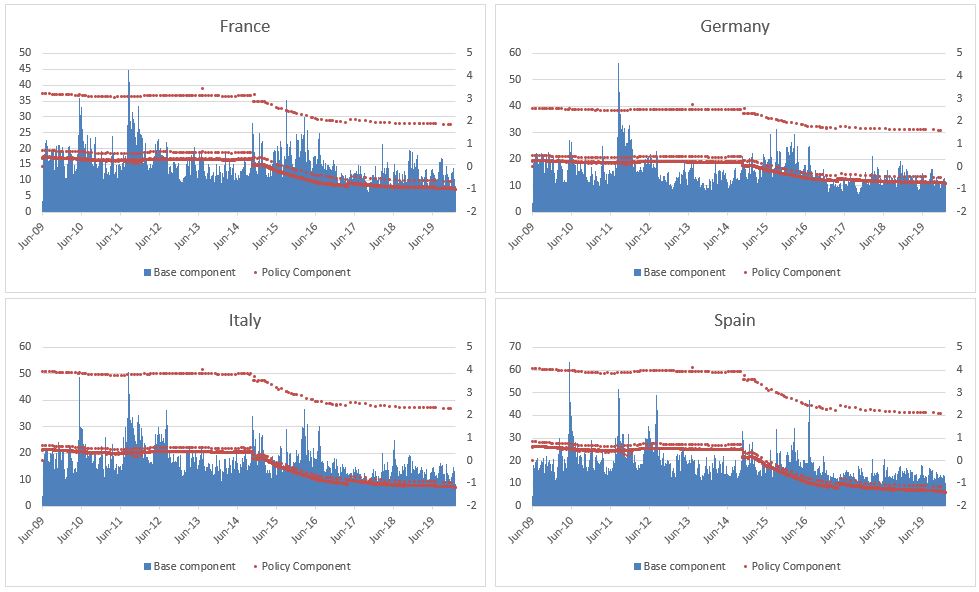}
	\caption{France, Germany, Italy and Spain Base (left axis) and Policy (right axis) component of volatility from the MAP. Sample period: June 1, 2009 to December 31, 2019} \label{fig:z_csi}
\end{figure}

This implementation effect can also be seen in Figure \ref{fig:z_csi}, which plots the evolution of the two volatility components (the blue line for the base component, left axis, and the red dotted-line, for the policy component, right axis\footnote{The line representing the $\xi$ process is defined in three different levels: the lowest depends on the characteristic path of the $\xi$ component; the highest represents, instead, volatility spikes due to the announcement effect; finally, the intermediate line, represents the level of volatility on days after a monetary policy announcement, when volatility comes back to its previous level.}). This effect is more evident starting from October 2014 - when the ECB implicitly communicated to the market that it would purchase corporate (next to government) bonds - coming in the form of a change in the slope of $\xi$ equation, lasting for the entire period of the program. This is one of the most interesting results, as it implies that the effectiveness of these policies rests on the total amount of securities held for monetary policy purposes, relative to ECB total asset. Moreover, an increase in this volatility component is observed in April 2017, which coincides with the reduction in the amount of securities purchased by ECB, set to \EUR 60 billion per month from the previous level of \EUR 80 billion. 

Results remain by and large the same, when we consider the Multiplicative MAP specifications. Once again, the proxies enter the models with the expected sign and with the highest level of significance. In both cases, coefficients are fairly lower, in view of the multiplicative way components combine: in other words, while in the MAP we generally have a negative $\xi_t$, within the multiplicative versions the effect by unconventional policies of dampening volatility is had when the $\xi_t$ process is less than 1 (by construction it is always positive): this requires lower proxies' coefficients, and this is achieved without imposing any kind of constraints. 

In order to compare economic effects across models, it is instructive to compute the corresponding marginal effects of the policy variables ($x_{t-1}$ and $\Delta_t$) on $\mu_t$: while in the additive specification they are constant and equal to the estimated coefficients, in the multiplicative specifications marginal effects are time varying. Looking at the average values\footnote{For what concerns $\Delta_t$, the marginal effects are considered only with respect to announcement days, and, as such, the average relates just to such days.} (Table \ref{tab:margina_effect}), the higher marginal effect of the implementation proxy is associated to the P-MAP. More specifically, a marginal increase in the proxy leads to a range of reductions in realized volatility comprised between -1.577 (Germany) and -3.382 (Spain); conversely, on announcement days volatility marginally increases, on average, by 2.603 and 4.146 points, in Germany and Spain, respectively. Overall, while unconventional policies had a higher impact on debt--troubled countries, the effect is present also for the others.

\begin{table}[t]
	\begin{center}
		\caption{Average marginal effects of policy variables on $\mu_t$ -- Sample period: June 1, 2009 - December 31, 2019 (parameter estimates by country as in Table \ref{tab:results}).}
		\vspace{2mm}
		\begin{tabular}{lrrr|rrr}
			& \multicolumn{3}{c|}{Marginal effect of $x_{t-1}$ on $\mu_t$ } & \multicolumn{3}{c}{Marginal effect of $\Delta_{t}$ on $\mu_t$} \\ [2mm]  
			& MAP        ~       & L-MAP    ~        & P-MAP      ~     & MAP         ~      & L-MAP    ~         & P-MAP     \\
			\hline
			France  & -1.836 ~ & -1.571 ~  & -2.172 ~ & 2.817 ~ & 2.236  ~ & 3.177  \\
			Germany & -1.328 ~ & -1.158 ~  & -1.577 ~ & 2.317 ~ & 1.821 ~ & 2.603  \\
			Italy   & -2.354 ~ & -2.033 ~  & -2.693 ~ & 3.448 ~ & 2.785 ~ & 3.953 \\
			Spain   & -2.823 ~ & -2.489 ~  & -3.382 ~ & 3.428 ~ & 2.912 ~ & 4.146         
		\end{tabular}\label{tab:margina_effect}
	\end{center}
    \vspace{-3mm}
	\begin{tablenotes}
		\footnotesize
		\item  \textit{Note:} the average marginal effect of $\Delta_t$ refers to the announcement days (see text).
	\end{tablenotes}
\end{table}

\begin{figure}[h]
\centering
\includegraphics[scale=0.65]{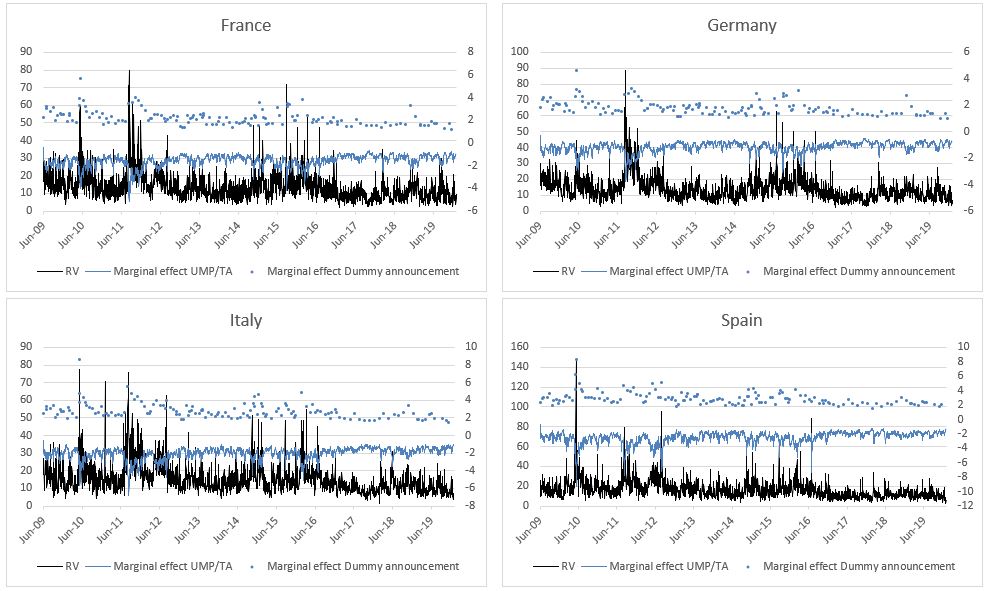}
\caption{France, Germany, Italy and Spain: marginal effects from the L-MAP. Sample period: June 1, 2009 to December 31, 2019 } \label{fig:marginal_effect_lacm}
\end{figure}

\begin{figure}[H]
\centering
\includegraphics[scale=0.65]{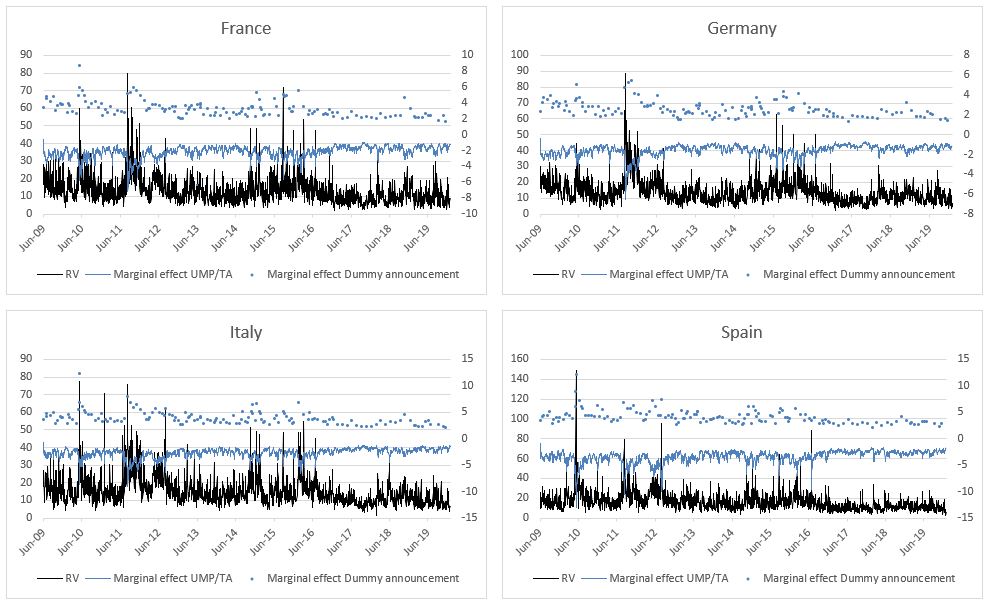}
\caption{France, Germany, Italy and Spain: marginal effects from the P-MAP. Sample period: June 1, 2009 to December 31, 2019} \label{fig:marginal_effect_liacm}
\end{figure}

The evolution of the marginal effects associated to the two multiplicative component specifications is shown in Figures \ref{fig:marginal_effect_lacm} and \ref{fig:marginal_effect_liacm}: for both models and for all the markets, marginal effects have a specular behaviour, to a certain extent mirroring the behavior of the realized volatility measure. In particular, the marginal effects of $\Delta_t$ have peaks in correspondence of volatility spikes, whereas the period of lower marginal effects of the proxy corresponds to periods of low volatility in the market. This result validates the dynamics in our models with the policy variables we adopted.


In the case of the MAP, the impact of unconventional policies on stock market volatility can also be measured as the ratio of the policy--related component $\xi_t$ to the general level of volatility $\mu_t$. The average of such a ratio signals a  reduction in volatility associated with the ECB unconventional policies between -0.6\% and -1.1\%, when measured for Germany and France, which becomes more marked for Italy (-1.24\%) and Spain (-1.38\%).  

As far as the dynamics of the $\xi_t$ component is concerned, the persistence effects, driven by the coefficient $\psi$, are fairly weak: such a coefficient is significant at 5$\%$ only for France across models and only marginally so for Germany for the L-MAP model. As a result, the data show that the policy--related volatility component follows the evolution of the policy proxy (with parameter $\delta$) and the changes occurring on announcement days (with parameter $\varphi$).

\FloatBarrier

\subsection{Out--of--sample Performance}
\label{sec:MCS}

In order to compare the one-step-ahead out-of-sample forecasting performance across the estimated models, we compute the Model Confidence Set  \citep[MCS,][]{Hansen:Lunde:Nason:2011} using the MSE and the QLike loss functions at the 10\% significance level.

For this purpose, we split the sample at the end of a year and we compute the one-step-ahead out-of-sample forecasts for the following year. The choice of the splitting dates is driven by the need to exclude the most important QE program, the EAPP, so that the first subsample considered stops at the end of 2014. We then consider each additional year in turn until 2018, in order to consider the full period of the EAPP. 

As shown in Table \ref{tab:mcs_res}, both loss functions provide similar results, confirming that the proxy--augmented models capture the features of the EAPP in delivering an improved forecasting performance. This outcome is more apparent when considering the subsamples individually. Before the EAPP announcement, the MCS results in the 2015 column of Table \ref{tab:mcs_res} show, for example, that all the models enter the best set of models (according to QLike, represented by the symbol $\Circle$) in 3 out of 4 cases, and even that the AMEM is the best model ($\CIRCLE$) in France, Germany and Spain and it belongs to the best set in Italy, for which the best model is the MAP (results are more mixed for the MSE).

Similar results derive from the forecasting period ending in 2016. For both loss functions, this time, the AMEM is the best model in 3 out of 4 cases, whereas in Spain such a role goes to the X-MAP: no component model is present in the best set in Germany, where the AMEM is the only model belonging to the best set (together with the P-MAP if we consider the MSE).

This good performance of the base AMEM is seriously  challenged by our models when we move to the forecasting period ending in 2017, characterized by a remarkable change in the ECB balance sheet composition: during this period the ECB purchased assets at a pace of \EUR 80 billion per month until March 2017 and \EUR60 billion per month until the end of the year. This time the policy--augmented models have a higher forecasting power and the AMEM is always excluded from the best set of models. Importantly, focusing on the QLike, the MAP enters the best set in 3 out of 4 cases, whereas the P-MAP is the best model for Germany and Spain. 

This latter model becomes the best across all cases considering the forecasting periods ending in 2018 and 2019, respectively. The difference in the results of these two sub-samples is in the best set of models, which is larger in the last year, possibly due to the fact that no APP was implemented until November 2019, when the ECB purchased new assets at a level of \EUR 20 billion per month. Overall, however, unconventional policies seem to have played a crucial role in reducing stock market volatility.

\begin{table}[H]
\begin{center}
\caption{The Model Confidence Set results. P-value 10\%}
\begin{tabular}{lrrrrr}
\multicolumn{6}{c}{France}                                                                                                                     \\
       & 2015                     & 2016                     & 2017                     & 2018                     & 2019                     \\
AMEM   & $\square$ $\CIRCLE$      & $\blacksquare$ $\CIRCLE$ &                          &                          &                          \\
X-MAP  &$\square$ {\color{white} $\square$}& $\square$ $\Circle$       & $\blacksquare$ $\CIRCLE$ &                          & $\square$ $\Circle$ \\
MAP    &                          & $\square$ $\Circle$       &      $\Circle$           &                          & $\square$  {\color{white} $\square$}              \\
L-MAP  & $\blacksquare$ $\Circle$   & $\square$ $\Circle$       &                          &                          & $\square$ {\color{white} $\square$} \\
P-MAP & $\square$ $\Circle$      & $\square$ $\Circle$       &  $\Circle$       & $\blacksquare$ $\CIRCLE$ & $\blacksquare$ $\CIRCLE$ \\
       &                          &                          &                          &                          &                          \\
\multicolumn{6}{c}{Germany}                                                                                                                     \\
       & 2015                     & 2016                     & 2017                     & 2018                     & 2019                     \\
AMEM   & $\blacksquare$ $\CIRCLE$ & $\blacksquare$ $\CIRCLE$ &                          & $\Circle$                 &                          \\
X-MAP  & $\square$ $\Circle$      &                          &                          &                          &                          \\
MAP    & $\square$ $\Circle$       &                          &                          &                          &                          \\
L-MAP  & $\square$ $\Circle$       &                          &                          &                          & $\square$ $\Circle$       \\
P-MAP & $\square$ $\Circle$      & $\square$  {\color{white} $\square$}              & $\blacksquare$ $\CIRCLE$ & $\blacksquare$ $\CIRCLE$ & $\blacksquare$ $\CIRCLE$ \\
       &                          &                          &                          &                          &                          \\
\multicolumn{6}{c}{Italy}                                                                                                                  \\
       & 2015                     & 2016                     & 2017                     & 2018                     & 2019                     \\
AMEM   & $\square$ $\Circle$      & $\blacksquare$ $\CIRCLE$ &                          &                          &                          \\
X-MAP  & $\square$ $\Circle$       & $\square$ $\Circle$       & $\blacksquare$ $\CIRCLE$ & $\blacksquare$      {\color{white} $\square$}     & $\square$ $\CIRCLE$      \\
MAP    & $\blacksquare$ $\CIRCLE$ & $\square$ $\Circle$       & $\Circle$                  &                          & $\square$ $\Circle$       \\
L-MAP  & $\square$ $\Circle$       & $\square$ $\Circle$       &                          &                          & $\square$ $\Circle$       \\
P-MAP & $\square$ $\Circle$       & $\square$ $\Circle$       &                          & $\square$ $\CIRCLE$      & $\blacksquare$ $\Circle$        \\
       &                          &                          &                          &                          &                          \\
\multicolumn{6}{c}{Spain}                                                                                                                    \\
       & 2015                     & 2016                     & 2017                     & 2018                     & 2019                     \\
AMEM   & $\CIRCLE$                & $\square$ $\Circle$       &                          &                          &                          \\
X-MAP  &$\Circle$              & $\blacksquare$ $\CIRCLE$ & $\blacksquare$ $\Circle$   & $\Circle$                  &                          \\
MAP    &                          & $\Circle$               & $\Circle$                 & $\Circle$                  &                          \\
L-MAP  &$\Circle$                & $\Circle$                  & $\Circle$                  &                          & $\square$   {\color{white} $\square$}             \\
P-MAP & $\blacksquare$ $\Circle$   & $\square$ $\Circle$       & $\square$ $\CIRCLE$      & $\blacksquare$ $\CIRCLE$ & $\blacksquare$ $\CIRCLE$
\end{tabular}\label{tab:mcs_res}
\end{center}
\begin{tablenotes}
      \footnotesize
      \item  \textit{Note:} $\square$($\Circle$) indicates models belonging to the best set according to the MSE(QLike) loss function; $\blacksquare$($\CIRCLE$) represents the best model.
    \end{tablenotes}
\end{table}

\subsection{Volatility responses to policy actions}
\label{sec:converge}

\begin{figure}[t]
	\centering
	\includegraphics[scale=0.6]{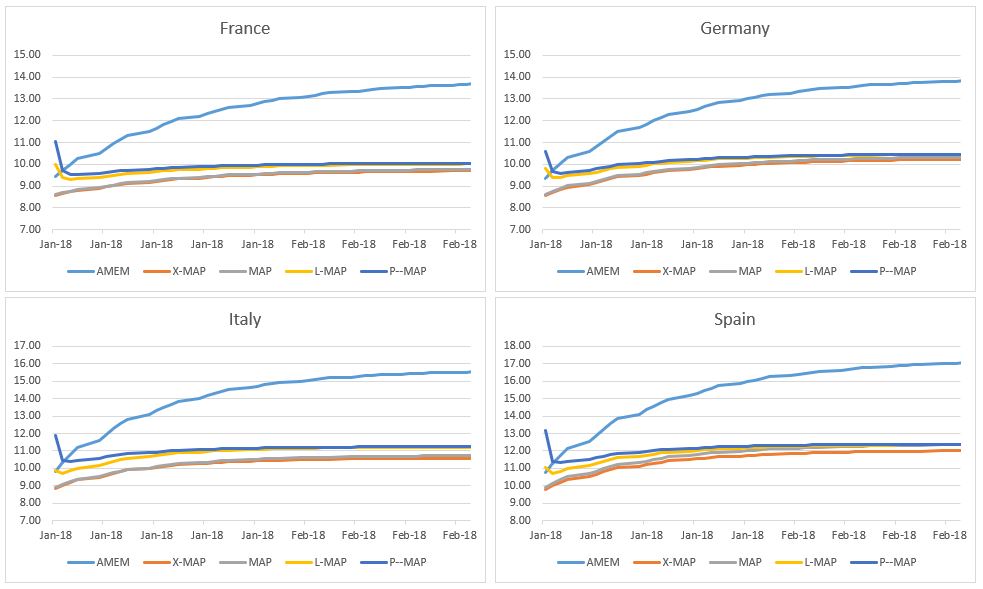}
	\caption{Multi-step forecast.  Sample period: June 1, 2009 to December 31, 2017. Forecast Period: January 1, 2018 - December 31, 2018} \label{fig:multistep2018}
\end{figure}
\FloatBarrier

In order to further analyze the dynamic effects of the policy actions on volatility, we graphically appraise the multi--step forecasts by model, as well as the profile of the impulse responses to a one standard deviation shock to the policy variable. For either case, we can evaluate the persistence of the effects and the time taken by different models in converging to the unconditional level of volatility.
This puts us in the position to address two substantive questions: i) for how long do QE policies affect stock market volatility? ii) what would be the volatility response to a higher QE shock?  

We focus on the sub-sample for the estimation period ending in 2017, with the following year used as the out--of--sample period and apply multi--step ahead forecasting,\footnote{The derivation of the formulas is available in the supplemental material.} obtaining the results shown in Figure \ref{fig:multistep2018}. As expected, the duration of the effect - measured as the number of days the volatility needs to reach the unconditional level, i.e. the unconditional mean\footnote{For practical purposes, convergence is considered achieved when subsequent forecasts do not differ by more than 1 basis point of volatility.} - depends on the country: we get an average duration between 41 business days (estimated in Italy via P-MAP) and 96 business days (from the MAP in Germany). Furthermore, regardless of the duration, once the unconventional policy effect is completely absorbed by the market, volatility converges to its unconditional level, which does not depend on the different model specifications; however, due to the different way they account for the policy effect, a slight difference between models presents itself, more marked for the AMEM, which neglects those effects altogether. As  expected, in all the cases the convergence path is upward sloping, with volatility reaching a higher level when the downward impact of unconventional policies ceases. Such an upward profile is shared by the AMEM as well, given that volatility is lower than its unconditional level at the beginning of the forecasting period.
 
\begin{figure}[t]
\centering
\includegraphics[scale=0.6]{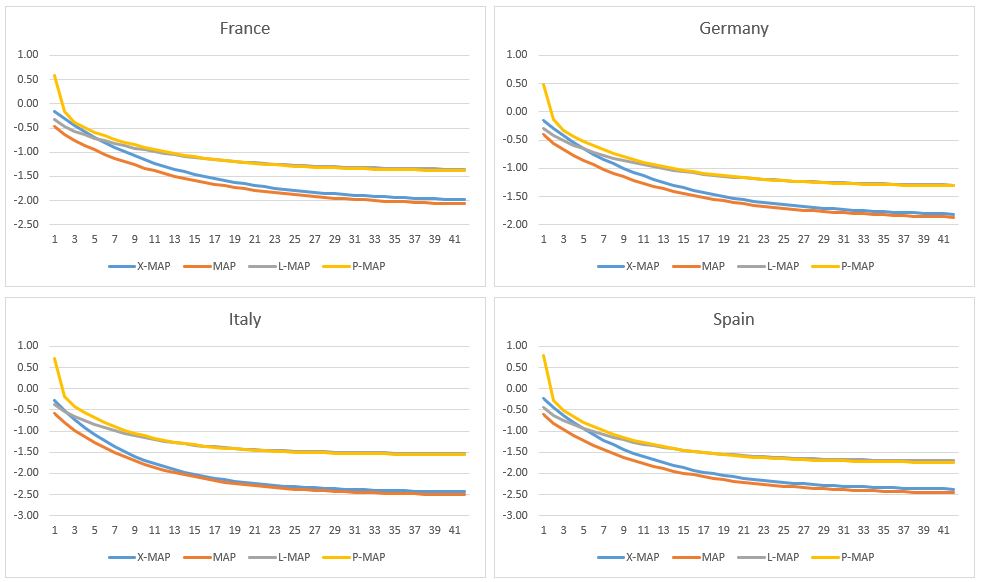}
\caption{Impulse Response Function: the effect on volatility of a 1 standard deviation unconventional policy shock. Sample period: June 1, 2009 to December 31, 2017. Forecast Period: January 1, 2018 - December 31, 2018} \label{fig:differences_2018}
\end{figure}

Finally, we analyze how volatility would have responded to a higher QE shock. For this purpose we have increased the QE proxies by one standard deviation (0.26). Taking the previous multi-step forecasting values as a baseline scenario,  in Figure \ref{fig:differences_2018} we represent the results as the difference between the two scenarios. The role played by unconventional policies in reducing stock  market volatility is clear: a higher unconventional policy shock would have had a stronger downward effect, with volatility that would have been lower down to 2.5 points in debt--troubled countries (2 points in France and Germany) relative to the baseline scenario. 

\FloatBarrier

\section{Concluding Remarks}
\label{sec:concl}

In this paper, we suggested a new class of models to analyze how unconventional monetary policies affect realized volatility. In detail, we suggested variants of multiplicative error models with two components, one representing what depends directly on QE policies, the other reproducing base volatility dynamics: such components may combine additively or multiplicatively. As shown by our results on ECB policies for four important Eurozone markets, what matters for the effectiveness of these policies is the balance sheet composition \citep[as argued by][]{Curdia:Woodford:2011}. Indeed, an increase in securities held by ECB for monetary policy purposes relative to total asset reduces volatility in both debt--troubled countries (Italy and Spain, in our analysis) and in benchmark countries (France and Germany), with more benefits for the former, in general. Admittedly, our policy proxies do not allow us to go deeper into the specific impact of each implemented policy, so that we cannot identify which of these extraordinary measures is more effective. This, of course, is an issue worth pursuing in future analysis, as well as the possibility to control also for spillovers among countries in a multivariate context, which could also highlight the presence of a common response across markets.

The presence of our proxy in the models entering the Model Confidence Set when evaluating the out--of--sample forecasting performance underlines the importance of the EAPP. This is also confirmed by the multi-step forecast procedure, given that, on the one hand, unconventional policies have lasting effects in lowering volatility for at most 90 business days, and, on the other, when shocking unconventional policies by one standard deviation stock market volatility decreases by up to 2.72 points. 

Exploiting the component structure of our models, we can extract a separate and distinctive signal related to the policy effects on volatility: between the additive and multiplicative way of combining the components, the results show a preference for the latter in estimation and out--of--sample, especially in 2018 and 2019. In economic terms, such results document that the unconventional monetary policy has a mitigating effect on market volatility at times of distress, when the interest rates are close to the zero lower bound, as a further tool available to central banks in order to restore the proper functioning of the economy.
\clearpage

\begin{center}
	{\large\bf SUPPLEMENTARY MATERIAL}
\end{center}
\begin{description}
	
	\item[Title:] Unconventional Policies Effects on Stock Market Volatility: A MAP Approach. Supplemental Material
	
	\item[Literature Review:] A summary of few empirical analysis concerning the effects of unconventional policies mainly on the bond market.
	
	\item[Marginal effects:] Derivation of marginal effects of the policy variables on volatility (Table \ref{tab:marginal_effects_formula}, Section \ref{sec:model} in the main text).
	
	\item[Multi-step forecasting:] Formulas used in the Multi--step forecasting procedure (Section \ref{sec:converge} in the main text) 
	
\end{description}

\bibliography{biblio}

\begin{thebibliography}{25}
\providecommand{\natexlab}[1]{#1}

\bibitem[{Andersen \textit{et~al.}(2006)Andersen, Bollerslev, Christoffersen
  and Diebold}]{Andersen:Bollerslev:Christoffersen:Diebold:2006}
Andersen, T.~G., Bollerslev, T., Christoffersen, P.~F. and Diebold, F.~X.
  (2006) Volatility and correlation forecasting, in \textit{Handbook of
  Economic Forecasting} (Eds.) G.~Elliott, C.~W.~J. Granger and A.~Timmermann,
  North Holland.

\bibitem[{Barndorff-Nielsen \textit{et~al.}(2008)Barndorff-Nielsen, Hansen,
  Lunde and Shephard}]{BarndorffNielsen:Hansen:Lunde:Shephard:2008}
Barndorff-Nielsen, O.~E., Hansen, P.~R., Lunde, A. and Shephard, N. (2008)
  Designing realised kernels to measure the ex-post variation of equity prices
  in the presence of noise, \textit{Econometrica}, \textbf{76}, 1481--1536.

\bibitem[{Bollerslev(1986)}]{Bollerslev:1986}
Bollerslev, T. (1986) Generalized autoregressive conditional
  heteroskedasticity, \textit{Journal of Econometrics}, \textbf{31}, 307--327.

\bibitem[{Bomfim(2003)}]{Bomfim:2003}
Bomfim, A.~N. (2003) Pre-announcement effects, news effects, and volatility:
  monetary policy and the stock market, \textit{Journal of Banking \& Finance},
  \textbf{27}, 133--151.

\bibitem[{Brownlees \textit{et~al.}(2011)Brownlees, Cipollini and
  Gallo}]{Brownlees:Cipollini:Gallo:2011}
Brownlees, C.~T., Cipollini, F. and Gallo, G.~M. (2011) Intra-daily volume
  modeling and prediction for algorithmic trading, \textit{Journal of Financial
  Econometrics}, \textbf{9}, 489--518.

\bibitem[{Brownlees \textit{et~al.}(2012)Brownlees, Cipollini and
  Gallo}]{Brownlees:Cipollini:Gallo:2012}
Brownlees, C.~T., Cipollini, F. and Gallo, G.~M. (2012) Multiplicative error
  models, in \textit{Volatility Models and Their Applications} (Eds.)
  L.~Bauwens, C.~Hafner and S.~Laurent, Wiley, pp. 223--247.

\bibitem[{Chan and Gray(2018)}]{Chan:Gray:2018}
Chan, K.~F. and Gray, P. (2018) Volatility jumps and macroeconomic news
  announcements, \textit{Journal of Futures Markets}, \textbf{38}, 881--897.

\bibitem[{Curdia and Woodford(2011)}]{Curdia:Woodford:2011}
Curdia, V. and Woodford, M. (2011) The central-bank balance sheet as an
  instrument of monetary policy, \textit{Journal of Monetary Economics},
  \textbf{58}, 54--79.

\bibitem[{D'Amico \textit{et~al.}(2012)D'Amico, English, L{\'o}pez-Salido and
  Nelson}]{dAmico:English:LopezSalido:Nelson:2012}
D'Amico, S., English, W., L{\'o}pez-Salido, D. and Nelson, E. (2012) The
  {F}ederal {R}eserve's large-scale asset purchase programmes: rationale and
  effects, \textit{The Economic Journal}, \textbf{122}, F415--F446.

\bibitem[{De~Santis(2020)}]{DeSantis:2020}
De~Santis, R.~A. (2020) Impact of the asset purchase programme on euro area
  government bond yields using market news, \textit{Economic Modelling},
  \textbf{86}, 192--209.

\bibitem[{Engle(1982)}]{Engle:1982}
Engle, R.~F. (1982) Autoregressive conditional heteroscedasticity with
  estimates of the variance of {U}nited {K}ingdom inflation,
  \textit{Econometrica}, \textbf{50}, 987--1007.

\bibitem[{Engle(2002)}]{Engle:2002}
Engle, R.~F. (2002) New frontiers for {ARCH} models, \textit{Journal of Applied
  Econometrics}, \textbf{17}, 425--446.

\bibitem[{Engle and Gallo(2006)}]{Engle:Gallo:2006}
Engle, R.~F. and Gallo, G.~M. (2006) A multiple indicators model for volatility
  using intra-daily data, \textit{Journal of Econometrics}, \textbf{131},
  3--27.

\bibitem[{Engle and Lee(1999)}]{Engle:Lee:1999}
Engle, R.~F. and Lee, G.~J. (1999) A permanent and transitory component model
  of stock return volatility, in \textit{Cointegration, Causality, and
  Forecasting: A Festschrift in Honor of Clive W. J. Granger} (Eds.) R.~F.
  Engle and H.~White, Oxford University Press, Oxford, pp. 475--497.

\bibitem[{Engle and Sokalska(2012)}]{Engle:Sokalska:2012}
Engle, R.~F. and Sokalska, M.~E. (2012) Forecasting intraday volatility in the
  {US} equity market. {M}ultiplicative component {GARCH}, \textit{Journal of
  Financial Econometrics}, \textbf{10}, 54--83.

\bibitem[{Eser and Schwaab(2016)}]{Eser:Schwaab:2016}
Eser, F. and Schwaab, B. (2016) Evaluating the impact of unconventional
  monetary policy measures: Empirical evidence from the {ECB}'s {S}ecurities
  {M}arkets {P}rogramme, \textit{Journal of Financial Economics}, \textbf{119},
  147--167.

\bibitem[{Georgiadis and Gr{\"a}b(2016)}]{Georgiadis:Graab:2016}
Georgiadis, G. and Gr{\"a}b, J. (2016) Global financial market impact of the
  announcement of the {ECB}'s asset purchase programme, \textit{Journal of
  Financial Stability}, \textbf{26}, 257--265.

\bibitem[{Ghysels \textit{et~al.}(2017)Ghysels, Idier, Manganelli and
  Vergote}]{Ghysels:Idier:Manganelli:Vergote:2017}
Ghysels, E., Idier, J., Manganelli, S. and Vergote, O. (2017) A high-frequency
  assessment of the {ECB} {S}ecurities {M}arkets {P}rogramme, \textit{Journal
  of the European Economic Association}, \textbf{15}, 218--243.

\bibitem[{Haitsma \textit{et~al.}(2016)Haitsma, Unalmis and
  de~Haan}]{Haitsma:Unalmis:DeHaan:2016}
Haitsma, R., Unalmis, D. and de~Haan, J. (2016) The impact of the {ECB}'s
  conventional and unconventional monetary policies on stock markets,
  \textit{Journal of Macroeconomics}, \textbf{48}, 101--116.

\bibitem[{Hansen \textit{et~al.}(2011)Hansen, Lunde and
  Nason}]{Hansen:Lunde:Nason:2011}
Hansen, P.~R., Lunde, A. and Nason, J.~M. (2011) The model confidence set,
  \textit{Econometrica}, \textbf{79}, 453--497.

\bibitem[{Otranto(2015)}]{Otranto:2015}
Otranto, E. (2015) Capturing the spillover effect with multiplicative error
  models, \textit{Communications in Statistics-Theory and Methods},
  \textbf{44}, 3173--3191.

\bibitem[{Patton(2011)}]{Patton:2011}
Patton, A.~J. (2011) Volatility forecast comparison using imperfect volatility
  proxies, \textit{Journal of Econometrics}, \textbf{160}, 246 -- 256.

\bibitem[{Shogbuyi and Steeley(2017)}]{Shogbuyi:Steeley:2017}
Shogbuyi, A. and Steeley, J.~M. (2017) The effect of quantitative easing on the
  variance and covariance of the {UK} and {US} equity markets,
  \textit{International Review of Financial Analysis}, \textbf{52}, 281--291.

\bibitem[{Toda and Yamamoto(1995)}]{Toda:Yamamoto:1995}
Toda, H.~Y. and Yamamoto, T. (1995) Statistical inference in vector
  autoregressions with possibly integrated processes, \textit{Journal of
  {E}conometrics}, \textbf{66}, 225--250.

\bibitem[{Voutsinas and Werner(2011)}]{Voutsinas:Werner:2011}
Voutsinas, K. and Werner, R.~A. (2011) New evidence on the effectiveness of
  ``{Q}uantitative {E}asing" in {J}apan, Tech. rep., CFS working paper.

\end{thebibliography}

\end{document}